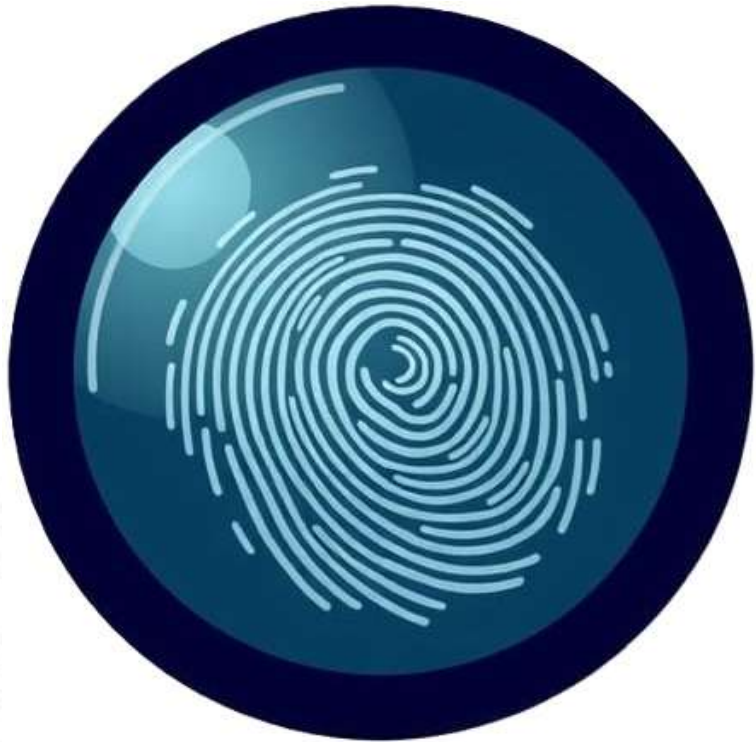

# The Birthmark Standard:

## Privacy-Preserving Photo Authentication
### via Hardware Roots of Trust and Consortium Blockchain

Technical Specification and Security Analysis
Version 1.1, February 2026

Sam Ryan

The Birthmark Standard Foundation

contact@birthmarkstandard.org

---

*Complete security analysis, privacy proofs, and implementation validation*

# 1 Abstract

The rapid advancement of generative AI systems has eroded the credibility landscape for photographic evidence. Modern image generation models produce photorealistic images indistinguishable from authentic photographs, undermining the evidentiary foundation upon which journalism and public discourse depend. Existing authentication approaches, such as the Coalition for Content Provenance and Authenticity (C2PA), embed cryptographically signed metadata directly into image files but suffer from two critical failures: technical vulnerability to metadata stripping during social media reprocessing, and structural dependency on corporate-controlled verification infrastructure where commercial incentives may conflict with the public interest. We present the Birthmark Standard, an authentication architecture leveraging manufacturing-unique sensor entropy derived from non-uniformity correction (NUC) maps to generate hardware-rooted authentication keys without requiring repeated sensor evaluation. During capture, cameras create anonymized authentication certificates proving sensor authenticity without exposing device identity via a key table architecture that maintains anonymity sets exceeding 1,000 devices. Authentication records are stored on a consortium blockchain operated by journalism organizations rather than commercial platforms, enabling verification that survives all forms of metadata loss. We formally verify privacy properties using ProVerif, proving observational equivalence for Manufacturer Non-Correlation and Blockchain Observer Non-Identification under Dolev–Yao adversary assumptions. The architecture is validated through a prototype implementation using Raspberry Pi 4 hardware with a camera sensor, demonstrating the cryptographic operations and authentication flow. Performance analysis based on cryptographic operation benchmarks and indexed database query characteristics projects <100ms camera overhead and <500ms verification latency at a scale of one million authentications per day.

# 2 Introduction

When Reuters pulled AI-generated images from their wire service in 2024, they identified the fakes using Adobe's Content Credentials [2,3]. While C2PA provides robust provenance metadata when preserved, it fails in the environment where misinformation actually spreads: social media platforms that strip authentication data through compression and format conversion [5,6,7]. Recent firmware vulnerabilities in camera C2PA implementations [23,24] demonstrate additional fragility. Even sophisticated cryptographic systems depend on secure implementation across the entire supply chain.

The structural problem runs deeper. Verification infrastructure is controlled by corporations whose commercial incentives may not prioritize the public good over their responsibility to shareholders. When engagement algorithms prioritize viral content over accurate content, platforms optimize for engagement. Camera manufacturers balance authentication features against production costs and competitive pressures. Journalists understand this risk. They've watched APIs change arbitrarily, seen platforms deprioritize news content, and experienced deplatforming for inconvenient reporting. Authentication infrastructure cannot depend on corporate goodwill.

When authentication infrastructure fails, whether through technical limitations or structural vulnerabilities, it creates a "liar's dividend" [4]: legitimate photographs can be dismissed as AI-generated simply because dismissal is now plausible. When visual evidence loses evidentiary weight, accountability mechanisms fail.

We present **the Birthmark Standard**, which addresses both technical and structural shortcomings through hardware roots of trust, metadata-independent blockchain storage, and consortium governance. The system uses camera sensor manufacturing variations (NUC maps) and smartphone PRNU patterns as physically unforgeable hardware fingerprints. During capture, devices create authentication certificates proving sensor authenticity, without exposing the device identity, through a key table architecture creating anonymity sets of thousands of devices. Authentication records are stored on a consortium blockchain operated by mission-aligned organizations rather than commercial platforms, enabling verification that survives all metadata loss.

Coalition governance matters. The blockchain is operated by journalism organizations, fact-checking networks, and press freedom advocates—"trust brokers" whose founding purpose centers on maintaining credible information infrastructure [31-33]. Organizations that produce media (Reuters, AP, BBC) have inherent editorial conflicts; organizations that validate, protect, or study media (WITNESS, Committee to Protect Journalists, International Fact-Checking Network, Reporters Without Borders) have aligned incentives without editorial bias. This isn't "blockchain for blockchain's sake." The system uses distributed ledger technology to create verification infrastructure that journalism organizations control rather than rent from corporations. The network continues operating even if individual members leave.

**Privacy by architectural separation.** Manufacturers can validate camera authenticity without seeing image content, location, or precise capture timing. Blockchain observers see cryptographic hashes without device identifiers. Submission servers cannot decrypt device tokens. Correlation requires compromising multiple independent systems simultaneously (manufacturers AND submission servers AND blockchain validators).

**Key technical contributions:**

1. **Chain-of-Custody Architecture:** A complete provenance pipeline from sensor capture through editing history to public verification, with security analysis demonstrating resistance to hardware compromise, coalition subversion, and privacy attacks.
2. **Dual Hardware Roots of Trust:** Sensor Non-Uniformity Correction (NUC) maps for professional cameras and PRNU-seeded keypairs for consumer smartphones prove physical sensor capture. These silicon fingerprints are manufacturing artifacts that AI generation cannot replicate without physical sensor access.
3. **Metadata-Independent Verification:** Authentication records stored on consortium blockchain while images remain with photographers. Verification operates on pixel data alone, surviving metadata stripping, format conversion, and all forms of social media processing.
4. **Privacy-Preserving Key Tables:** Cryptographic architecture enabling manufacturer validation of hardware authenticity while maintaining k-anonymity ($k \approx 1{,}000$ devices per anonymity set). No single entity can correlate specific photographer identity to image registry without compromising multiple independent systems simultaneously.
5. **Consortium Blockchain Governance:** Blockchain operated as public trust by mission-aligned organizations rather than corporate entities. Decentralized governance prevents single-point control without cryptocurrency exposure or speculative token economics.
6. **Tiered Storage Protocol (Hot/Cold Nodes):** Scalable architecture where permanent "Cold" nodes ensure long-term archival integrity while high-performance "Hot" nodes prune historical state, maintaining lean ledger footprint of approximately 150 bytes per authentication record.
7. **Optimized Performance & Caching:** Camera-side overhead below 100ms, with distributed caching protocol enabling social media platforms to render authenticity badges with sub-millisecond latency at scale.

**Scope and limitations.** This system verifies that images originated from legitimate camera hardware rather than AI generation. However, it does not verify scene authenticity. Staged photographs pass validation. Authentication proves "this came from a camera, not Midjourney," not "this depicts events as claimed."

**Implementation as public infrastructure.** This work is structured as a 501(c)(3) nonprofit foundation because authentication infrastructure serving public discourse must be operated by organizations whose mission aligns with truth rather than profit. Anyone can verify image authenticity by hashing an image locally and querying the blockchain directly, requiring no subscription, authentication credentials, or platform intermediary. All specifications are published under Apache 2.0 and AGPL-3.0 licenses as defensive prior art, preventing patent enclosure and enabling manufacturer adoption without licensing fees. If this Foundation fails, the work survives.

We validate the architecture through prototype implementation using a Raspberry Pi 4 with a camera sensor, demonstrating the complete cryptographic pipeline from photon capture to blockchain verification.

The remainder of this paper is organized as follows. Section 2 surveys related work. Section 3 presents our threat model and design goals. Section 4 describes the complete system architecture. Section 5 provides security analysis. Section 6 presents implementation details and performance characteristics. Section 7 discusses adoption pathways, governance framework, and relationship to existing standards. Section 8 concludes with impact assessment and future directions.

# 3 RELATED WORK

**Content Provenance:** The Coalition for Content Provenance and Authenticity (C2PA) embeds cryptographically signed manifests into image files, providing rich provenance when metadata preserved. However, social media platforms systematically strip metadata through compression and format conversion, affecting the vast majority of shared images. Firmware vulnerabilities further compromise trust in embedded credentials.

**Camera Authentication:** Photo Response Non-Uniformity (PRNU) analysis exploits pixel-level sensitivity variations for forensic source identification. Non-Uniformity Correction (NUC) maps provide manufacturing-calibrated sensor fingerprints. We use NUC maps and PRNU analysis as hardware roots of trust for real-time authentication rather than post-hoc forensic analysis.

**Blockchain Systems:** Bitcoin demonstrates foundational timestamping but creates privacy concerns in permissionless designs. Consortium frameworks like Hyperledger Fabric provide controlled validators but lack built-in governance. We use Substrate for forkless runtime upgrades through on-chain governance, addressing operational burden where traditional blockchains require significant coordination overhead per upgrade.

**Privacy-Preserving Authentication:** Zero-knowledge proofs enable authentication without revealing identity but require complex computations unsuitable for battery-powered cameras with millisecond response requirements. We achieve privacy through architectural separation (encrypted tokens + key table anonymity sets), enabling lightweight camera-side operations compatible with mobile deployment constraints.

**Positioning:** We provide blockchain-based authentication surviving metadata stripping (addressing the high failure rate of metadata dependent solutions), hardware roots of trust defending against AI generation (stronger than software-only approaches), and privacy-preserving architecture enabling manufacturer validation without surveillance capability (balancing authentication integrity with photographer privacy).

# 4 THREAT MODEL AND DESIGN GOALS

We formalize security properties required for trustworthy photographic authentication by specifying adversary capabilities and explicit design goals.

## 4.1 Threat Model

**Compromised Infrastructure:**

Adversaries may compromise submission servers or validator nodes to manipulate authentication records, selectively censor images, or extract data. This includes malicious operators, compromised credentials, or software vulnerabilities. We assume adversaries can compromise up to 33% of validator nodes (standard Byzantine threshold). Compromised servers can inspect submitted data but cannot directly modify blockchain state without validator cooperation. Submission servers cannot identify specific devices (only encrypted tokens visible) or photographers.

**Hardware Attackers:**

Adversaries attempt to extract secrets from camera secure elements using reverse engineering tools including focused ion beam microscopy and power analysis equipment. We assume Secure Element certification (EAL5+) provides reasonable protection with per-camera compromise costs exceeding $100,000. Successful extraction affects cameras sharing compromised key tables.

**Privacy Adversaries:**

Adversaries seek to correlate photographers across images or track devices through statistical analysis of authentication patterns and network metadata. This includes surveillance of manufacturer validation traffic, blockchain analysis, and network timing attacks. Goal is to link photographers to their images or track device usage patterns.

**Global Passive Surveillance:**

Nation-state actors with capabilities including traffic analysis, infrastructure compromise, and long-term data collection attempt to track photographers through authentication metadata. Cannot violate physical constraints (forging sensor characteristics) or cryptographic assumptions (breaking AES-256, SHA-256). May attempt to compel infrastructure operators but cannot force compliance without detection.

**Out of scope:** We do not protect against staged photographs (authentication verifies hardware, not scene truthfulness).

## 4.2 Design Goals

**G1: Hardware-Rooted Authenticity.** The system must cryptographically prove images originated from legitimate camera hardware rather than AI generation. Adversaries cannot generate valid authentication tokens without possessing legitimate hardware or successfully compromising hardware security mechanisms (computationally infeasible).

**G2: Privacy Preservation.** The system provides computational privacy for photographer activity through separation of knowledge across system components.

**Formal Definition:** The system achieves three privacy properties:

(1) Manufacturer Non-Correlation: Manufacturers learn device identity (device_id) but have zero knowledge of authenticated outputs. Formally, given camera $C_i$ authenticating images $I_1, I_2, ..., I_n$, manufacturer M observes authentication events but P(M correlates any (device_id, image_hash) pair) = 0, as image hashes are never transmitted to M during validation.

(2) Blockchain Observer Non-Identification: Blockchain observers see image hashes and obscured processing timestamps but cannot identify devices, manufacturers, or photographers. The Authentication records contain no manufacturer identifiers. Given authentication record R containing only image_hash and timestamp, observer O cannot determine which device or manufacturer created R (k-anonymity across all authenticated devices).

(3) Submission Server Blindness: Submission servers see encrypted tokens but cannot identify devices or photographers. Given encrypted token T from camera $C_i$. server S cannot decrypt T to learn device_id (only manufacturer M possesses decryption keys).

**Adversary Model:** Privacy guarantees hold against computationally-bounded adversaries who cannot break AES-256 encryption or SHA-256 collision resistance. Manufacturer with access to validation keys can identify devices but not correlate to image content (separation of validation and content channels). Coalition of blockchain validators cannot identify devices (device IDs never recorded on chain or viewed unencrypted in the content channel).

This provides computational privacy rather than information-theoretic privacy. Adversaries with sufficient resources to compromise multiple system components simultaneously could correlate activity, but this requires compromise of both manufacturer key infrastructure AND blockchain validation infrastructure, exceeding our threat model's Byzantine threshold (33%).

**G3: Metadata Independence.** The Authentication records must survive all metadata loss including format conversion, compression, cropping, and platform reprocessing. Verification requires only pixel data.

**G4: Decentralized Trust.** No single entity can unilaterally control authentication decisions or manipulate records. Coalition governance requires supermajority (67%) for major decisions. Legal compulsion results in validator removal, not network compromise.

**G5: Public Verifiability.** Anyone must be able to verify image authenticity without subscription, authentication, or trusted intermediary. Verification process cannot reveal verifier identity.

**G6: Scalability.** Architecture must support 1 million+ authentications per day with <100ms camera overhead, <500ms verification latency, and blockchain storage growth under 100GB/year at ~$100/month per node. Design must scale to global adoption (~5 billion/day), though infrastructure costs at that scale would require expanded funding.

**G7: Graceful Degradation.** Authentication failures cannot prevent photography. Cameras continue normal operation with authentication temporarily disabled during Secure Element failures, network unavailability, or submission server unreachability. Cameras queue authentication records during outages and process them when systems restore, ensuring no permanent loss of authentication capability.

# 5 SYSTEM ARCHITECTURE

The Birthmark Standard establishes a complete chain of custody from photon capture through public verification. The architecture has six integrated components operating with separated concerns for privacy preservation.

## 5.1 System Overview and Data Flow

The authentication pipeline transforms camera-generated cryptographic proofs into compact blockchain records through a multi-stage validation process. This section presents the complete data structures transmitted between system components.

**Camera Submission Packet**

When a camera captures an image, it generates a complete authentication packet containing three components: the birthmark record (image identifiers and optional metadata hashes), the manufacturer certificate (cryptographic proof of camera authenticity), and a camera signature binding these elements together. The camera transmits this packet to a submission server for validation and blockchain posting. Privacy is ensured through data separation as demonstrated below.

The birthmark record contains the core image data that will ultimately be stored on the blockchain:

- **Image Hash:** SHA-256 hash of the pixel data
- **Modification Level:** Indicates (0) raw sensor data, (1) validated content, or (2) content modified
- **Parent Image Hash:** Links to the authenticated parent image for crops or edits; null for original captures (0 or 32 bytes)
- **Optional Metadata:** If the photographer has enabled metadata recording, the packet includes HMAC-SHA256 hashes of timestamp, geolocation, and owner identity (24 bytes total). These hashes enable verification by anyone with the original image while preserving privacy on the public blockchain.

The manufacturer certificate provides cryptographic proof that the image originated from a legitimate camera:

- **Validator ID:** Routes the certificate to the appropriate manufacturer validator (e.g., "CANON_001")
- **Encrypted Token:** The camera's NUC hash encrypted with AES-256-GCM, enabling manufacturer validation without exposing device identity to submission servers
- **Key Reference:** Identifies which encryption key table and index the manufacturer should use for decryption, creating anonymity sets of >1,000 cameras

- **Deviation Test Data:** For ISP-processed or edited images, includes the list of operations performed, their adjustment values, the cumulative deviation score, and a hash of the calculation code used

- **Record Hash:** SHA-256 hash of the entire birthmark record, binding the certificate to specific image data

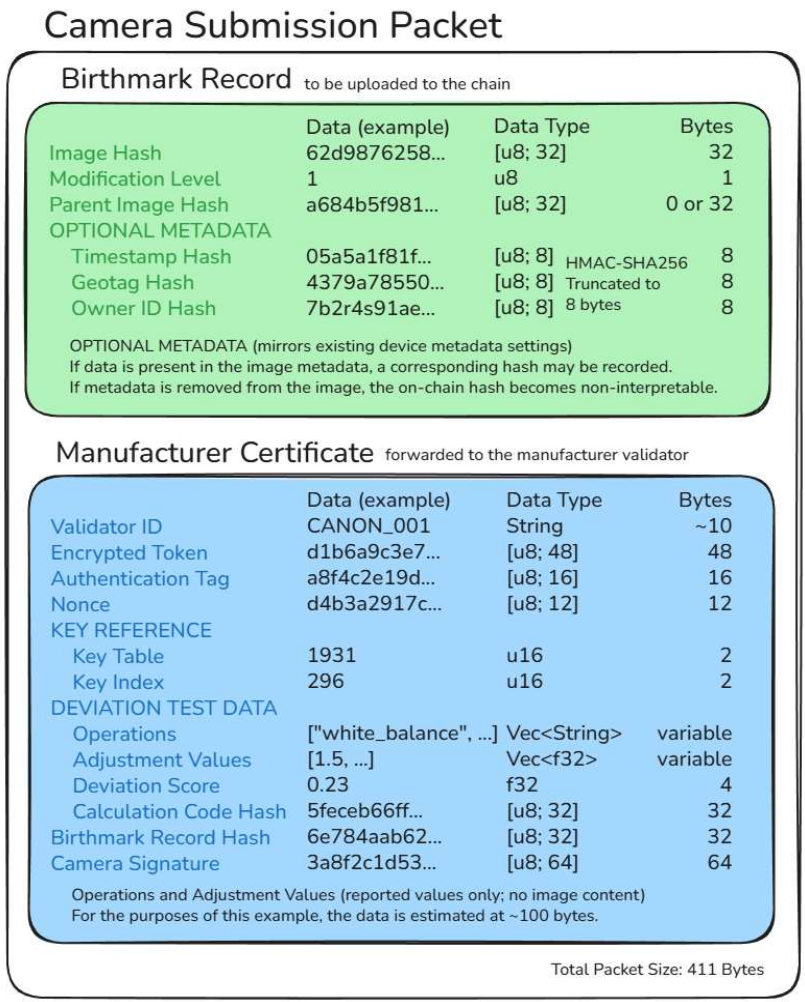

The camera signature (64 bytes, ECDSA secp256k1) signs the manufacturer certificate, cryptographically proving that this specific camera authorized the authentication request for this specific image. The complete packet totals approximately 411 bytes.

The submission server receives this packet and performs initial validation: verifying the camera signature matches the certificate, confirming the record hash binds the certificate to the birthmark data, and forwarding the manufacturer certificate to the appropriate validator for camera authentication.

*Figure 1: Camera → Submission Server. Complete authentication packet (~411 bytes) showing birthmark record (green), manufacturer certificate (blue), and camera signature.*

**Blockchain Submission and Storage**

Once hardware authenticity is verified, the validator checks the proposed Modification Level based on declared software tools in the Deviation Test Data, using stochastic audits to flag discrepancies between reported operations and pixel data (see 5.4). The resulting integrity classification is then bundled with the hardware proof and transmitted to a submission server which submits validation proofs to blockchain validators.

The blockchain validators hold onto the validation data until two matching records are received (dual-approval protocol) proving that two independent submission servers and the manufacturer validator all approved this authentication. From each server:

- **Server ID:** Generic node identifier (e.g., "node-eu-1") indicating geographic location without revealing operator identity (~10 bytes)

- **Server Signature:** ECDSA signature by that server over the birthmark record hash, proving independent validation (64 bytes)

- **Birthmark Record Hash:** The same hash from the manufacturer certificate, binding all signatures to the same record (32 bytes)

- **Validator Signature:** The manufacturer's ECDSA signature approving this specific record hash, proving camera authenticity (64 bytes)

Both servers include the same birthmark record hash and validator signature, proving they validated the same camera submission and received the same manufacturer approval. Blockchain validators verify all signatures (camera, manufacturer, and both servers) before accepting the record. This validation data totals approximately 350 bytes but is discarded after verification—it serves as proof for validators but need not be stored permanently.

The blockchain record contains only the essential data required for long-term verification:

- **Image Hash, Modification Level, Parent Hash, Metadata Hashes:** Copied from the camera's birthmark record
- **Posting Server IDs:** Concatenated identifiers of both validating servers, proving dual-approval (e.g., "node-eu-1/node-us-3")
- **Posting Timestamp:** Server processing time rounded to 10-minute intervals, providing temporal context while preserving privacy

*Figure 2: Submission Server → Blockchain Validators. Validation data (~350 bytes, pink) verified then discarded; only birthmark record (153 bytes, green) stored permanently on-chain.*

This data transformation—from complete cryptographic proof (411 bytes) through multi-party validation (350 bytes of signatures) to compact permanent record (153 bytes)—enables the system to maintain strong security properties while scaling to millions of daily authentications. At 1 million authentications per day, annual blockchain storage grows by only 56 GB, keeping validator node costs accessible to nonprofit journalism organizations.

## 5.2 Hardware Root of Trust

**Manufacturing-Calibrated Cameras**

Professional cameras undergo per-pixel sensitivity calibration during manufacturing, producing Non-Uniformity Correction (NUC) maps that compensate for photodiode response variations. NUC_hash = SHA-256(NUC_map) serves as an unforgeable hardware fingerprint due to quantum-level manufacturing variations. Manufacturers maintain secure databases containing NUC hashes from legitimate cameras, enabling device authentication without storing device identifiers. The database answers only membership queries: 'Is this NUC hash from a legitimate camera?' without revealing which specific camera. Cameras store ECDSA keypairs in hardware secure elements (TPM 2.0, ARM TrustZone) with private keys that never leave the secure boundary.

**Smartphone Implementation**

Consumer smartphones lack manufacturing-time calibration but possess Photo Response Non-Uniformity (PRNU) - stable sensor noise patterns from silicon imperfections. During first launch, the app captures 20 images, extracts PRNU pattern, uses it as entropy for ECDSA keypair generation, stores the private key in platform secure element (Android StrongBox, iOS Secure Enclave), then permanently discards the PRNU pattern. The keypair becomes permanent device identity. Factory reset destroys authentication capability permanently—the same hardware can re-register but generates a different keypair, appearing as a new camera to the system. This approach trades re-registration capability for hardware-rooted entropy without requiring manufacturing-time calibration.

**Metadata Privacy**

Timestamp hashes use month-level precision ("2025-11") to limit operational tempo tracking. Geolocation and owner hashes employ HMAC-SHA256(data, random_nonce) truncated to 64 bits, with nonces stored in image EXIF. The 2^64 search space resists brute-force even with known approximate locations. Domain separation ("BM-v1-timestamp:", "BM-v1-geolocation:", "BM-v1-owner:") prevents cross-field attacks if one field is compromised.

## 5.3 Privacy-Preserving Key Architecture

Key tables enable manufacturer authentication without revealing device identity to submission servers. Each camera is randomly assigned to multiple tables from a global pool, with each table containing hundreds to thousands of AES-256 keys. This creates anonymity sets where each key corresponds to thousands of cameras. During submission, the camera includes key_table_id and key_index, enabling manufacturer validation without identifying the specific device. Manufacturers observe encrypted tokens and validation requests but cannot correlate them to specific blockchain image hashes without compromising submission servers.

Keys remain valid indefinitely unless compromise is detected through validation pattern anomalies (e.g., single key validating statistically improbable image volumes) or security audits. Upon detection, manufacturers issue targeted key replacements via OTA update to affected devices and immediately revoke compromised keys from validator databases. The large anonymity set size and distributed table architecture limit the attack surface from individual key compromises.

This event-driven approach minimizes operational overhead while enabling rapid response to actual security incidents. Manufacturers may optionally implement periodic rotation aligned with firmware releases if security policies require it, but scheduled rotation is not protocol-mandated. The specific number of tables, keys per table, and resulting anonymity set sizes are deployment parameters chosen by each manufacturer based on their device fleet size and privacy requirements.

**Manufacturer validation is stateless:** MA confirms NUC hash membership in legitimate device set but does not log device identifiers. Even MA database compromise reveals only the population of legitimate devices, not which devices validated which images or when specific cameras were active.

**Device Revocation:** When fraud is detected (stolen credentials, forged authentications), the compromised NUC hash fingerprint is added to the MA revocation list. This blocks future validations using that specific fingerprint but does not prevent the physical camera from taking pictures, only from interacting with the Birthmark system. Revocation targets fraudulent credentials, not hardware. If supported by the manufacturer, devices can be re-registered using PRNU measurements based on the standard for smartphone devices, generating different fingerprints. Reprovisioning is subject to a manufacturer-defined security threshold, typically requiring a hardware-rooted attestation to prove the physical sensor remains under the control of the original secure element.

**Small Manufacturer Aggregation:** Manufacturers with limited production volumes (<1,000 cameras annually) may achieve insufficient anonymity set sizes independently. Birthmark Systems LLC offers commercial MA infrastructure that pools cameras across multiple small manufacturers into shared key tables. For example, Fairphone (500 cameras/year), an action camera manufacturer (300 cameras/year), and a documentary equipment company (200 cameras/year) would share validation infrastructure, achieving k ≥ 1,000 through aggregation. Validators see only the commercial MA identifier (e.g., "BIRTHMARK_MA_001"), not individual manufacturer identities, improving privacy beyond what small manufacturers could achieve independently while reducing operational burden.

## 5.4 Software Tool Use Reporting and Integrity Audit

The system enforces editorial policy through declared tool usage combined with non-deterministic integrity audits, providing defense-in-depth beyond cryptographic authentication.

**Policy Layer - Software Tool Use Reporting**

Editing software declares operations performed on authenticated images (exposure adjustments, white balance, denoise, crop), defining modification classifications:

- Level 0 (Unprocessed): Raw sensor data
- Level 1 (Tonal/Compositional): Exposure ±2 stops, white balance, denoise, crop
- Level 2 (Content Modification): Clone stamp, object removal, generative fill

News organizations set policies: "Only accept Level 0-1 for front-page stories" or "Flag Level 2 for editor review." Software declarations enable automated policy enforcement.

**Integrity Audit - Non-Deterministic Deviation Testing**

Software Authority performs non-deterministic audits verifying software performed only declared operations. The audit extracts random image patches from authenticated parents, simulates declared operations, measures deviation between simulated and actual results. High deviation indicates undeclared modifications.

**Critical distinction:** Deviation measures difference from EXPECTED results given reported operations. Lightroom applying "Exposure +1.5 stops" produces low deviation (~8-12%) because Software Authority accurately replicates that adjustment. Content alteration tools cannot simulate expected results—modifications depend on semantic understanding not captured in operation parameters—producing deviation >40%.

**Information Asymmetry and Adversarial Resistance**

Audit methodology remains nondeterministic: sampling locations, patch sizes, thresholds, and statistical models evolve without public disclosure. This information asymmetry prevents adversarial optimization—developers cannot tune manipulation to pass audits without knowing detection parameters.

Suspicious patterns trigger red flag clusters (multiple suspicious validations from same software/manufacturer/timeframe). Coalition analysts investigate clusters, potentially blacklisting compromised versions. Analogous to financial fraud detection: occasional anomalies are noise, systematic patterns indicate coordinated attacks.

**Defense-in-Depth Architecture**

This audit layer complements cryptographic authentication without creating dependencies. Compromising deviation testing does not compromise core authentication (hardware roots of trust, manufacturer validation, blockchain integrity, privacy architecture). The audit layer increases attack cost and detection probability without creating single points of failure.

**Scope and Validation**

Rigorous validation requires proprietary ISP pipelines, editing software internals, and large labeled datasets—resources requiring manufacturer partnerships and funding. Initial deployment uses conservative thresholds based on known ISP processing characteristics. Threshold tuning improves iteratively through coalition feedback and empirical testing, positioning integrity auditing as an evolving defense mechanism rather than fixed cryptographic guarantee.

## 5.5 Submission Security Architecture

Three-layer defense prevents packet tampering, server forgery, and pattern correlation:

**Camera Signature Binding:** Camera creates authentication record, hashes it, creates certificate {encrypted_token, token_excerpt, record_hash}, signs with SE private key. Server validates: device public key certificate (Manufacturer Authority signed), cert signature (camera signed), record matches

cert.record_hash, token binds via excerpt. Prevents image hash swap, cert modification, token reuse, component mixing.

**Dual-Approval Protocol:** Blockchain requires two MA-signed approvals from different submission servers. Camera submits to two servers simultaneously. Each validates independently, obtains Manufacturer Authority approval. Blockchain enforces: two approvals required, different server_ids, both signatures verify with Manufacturer Authority public key. Single compromised server cannot forge second approval (Manufacturer Authority signatures bind to server_id). Prevents single-server fraud.

**Server Selection and Rotation:** Every 24 hours, cameras update their server selection set by querying the server registry (https://registry.birthmarkstandard.org/servers.json) and selecting three geographically diverse servers (e.g., Americas, Europe, Asia-Pacific) excluding top 25% busiest. Submissions distributed round-robin. Privacy: pattern fragmentation (no single server sees complete timeline), geographic resistance (multi-jurisdictional surveillance required), temporal breaking (daily rotation prevents long-term tracking).

## 5.6: Design Philosophy: Architectural Tradeoffs

Every authentication system must balance three requirements: Effectiveness (preventing spoofing, surviving real-world distribution, scaling to billions of images), Privacy (preventing surveillance, protecting photographer anonymity), and Simplicity (minimizing architectural complexity and operational overhead). These form a constraint triangle where optimizing any two forces tradeoffs in the third.

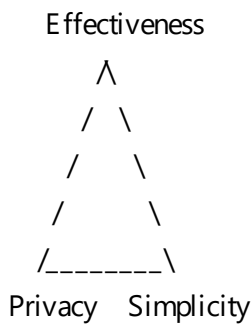

```
    Effectiveness
         /\
        /  \
       /    \
      /      \
     /________\
  Privacy   Simplicity
```

Real-world implementations demonstrate all three tradeoff choices and their consequences.

**Watermarks** (simple + private): News organizations used these for decades. They get cropped out, removed in compression, and offer no cryptographic proof of authenticity—anyone can add a watermark to any image.

**C2PA** (effective + simple): Strong cryptographic signatures with straightforward implementation, but centralized corporate consortium governance. Works effectively when metadata is preserved.

**Birthmark Standard** (effective + private): Images stay verifiable after metadata stripping. Journalists can't be de-anonymized through the authentication system. Designing for these outcomes requires architectural complexity (e.g., coalition governance, jurisdictional separation, cross-system encryption, distributed validation).

Architectural complexity doesn't mean complexity for the user. The system is configured to run automatically is designed to be invisible. Signal has complex cryptography but works like SMS. Tor uses multi-hop routing but browses with one click.

This choice, effectiveness and privacy over architectural simplicity, drives the design decisions in the following sections.

## 5.7 Consortium Blockchain Registry

GRANDPA consensus (67% validator agreement) ensures single canonical record set. Authentication records contain: image_hash (SHA-256 of pixels), modification_level (0=raw sensor data, 1=validated content, 2=content modified), parent_image_hash (links to authenticated parent for crops/edits), optional metadata hashes (timestamp, geolocation, owner), posting_server_ids (identifiers of both validating servers), and posting_timestamp (processing time rounded to 10-minute intervals). Records are immutable and append-only; edited images create new records linking to parent via parent_image_hash, forming verifiable custody chains.

**Validator Management:** New validators join via 67% governance approval, $5-10K purchase fee, technical setup. Exit voluntary (notice period) or involuntary (emergency vote if compromised/compelled). Byzantine tolerance maintains security during transitions (minimum 4 validators tolerates 1 faulty, recommended 10 tolerates 3 faulty). Trust broker stability (journalism organizations, fact-checking networks, press freedom advocates with decades-long operations) suggests low attrition.

**Storage:** 153 bytes per record. At 1 million images/day: 56GB/year blockchain growth. Validators maintain hot storage (SSD), cold archive (HDD), distributed backups. Query performance: O(1) hash lookup via indexed database.

## 5.8 Public Verification

Anyone can verify image authenticity. Compute SHA-256(pixel_data), query any validator node via JSON-RPC, receive authentication record if exists. No subscription, authentication, or platform intermediary required. Verification survives all metadata stripping, format conversion, compression. Browser extensions can check images automatically; social platforms can integrate verification badges.

**Anti-Correlation Design:** Device identifiers never appear on blockchain records. Manufacturers observe device identities during validation but cannot link them to specific image hashes without compromising submission servers. Submission servers observe image hashes but cannot decrypt device identities. Correlation requires compromising both manufacturer validation infrastructure AND submission server logs simultaneously, exceeding the system's Byzantine fault tolerance threshold (33% of validators).

# 6 FORMAL SECURITY ANALYSIS

This section provides formal analysis of security properties, adversary capabilities, and defense mechanisms. We state security goals precisely, define adversary models, present security claims with proof sketches, and acknowledge limitations.

## 6.1 Security Goals

The Birthmark Standard achieves four primary security goals:

**Authentication Integrity:** Only legitimate cameras (possessing a valid NUC map hash or PRNU-seeded key pair) can create blockchain-accepted authentication records. Adversaries cannot forge records without compromising hardware roots of trust or manufacturer authority cryptographic keys.

**Privacy Preservation:** Manufacturer authorities cannot correlate specific devices to specific images even if the system experiences cataclysmic compromise. Blockchain observers cannot identify which camera created which image. Submission servers cannot determine device identity from encrypted tokens. These properties hold given standard cryptographic assumptions (AES-256 security, k-anonymity sets).

**Byzantine Fault Tolerance:** System maintains authentication integrity and consensus with up to $\lfloor (n-1)/3 \rfloor$ compromised validators where n = total validator count. Malicious validators cannot insert fraudulent records or block legitimate submissions without exceeding Byzantine threshold.

**Anti-Censorship:** No single entity (submission server, validator, manufacturer) can selectively censor specific photographers or images. Removal requires either dual-server collusion or validator majority consensus, both exceeding Byzantine threshold assumptions.

**Verification Approach:** We provide security analysis using applied π-calculus and ProVerif verification for core privacy properties (Appendix A). Additional properties are analyzed via proof sketches demonstrating how architectural separation prevents correlation attacks. While mechanized formal verification across all properties using proof assistants remains future work, our approach follows standard practice for deployed cryptographic systems where ProVerif validation establishes confidence in protocol design. Most production security systems (TLS, Signal Protocol, WireGuard) rely on similar combinations of mechanized proof for critical properties and rigorous argument for secondary properties rather than complete mechanized verification.

## 6.2 Adversary Capabilities

We model three adversary types with distinct capabilities:

**Hardware Adversary (A_HW):** Physical access to camera devices. Can extract secure elements via invasive techniques (cost: ~$100K per device, affects single unit). Can attempt to forge NUC maps or PRNU patterns (requires manufacturing access or cryptographic break). Cannot compromise random number generators if PRNU-seeded provisioning used (physical entropy source).

**Network Adversary (A_NET):** Controls network infrastructure. Can compromise up to 33% of validator nodes (Byzantine threshold assumption). Can compromise submission servers (but dual-approval requires two independent servers). Observes all network traffic including timing, packet sizes, frequency patterns. Can perform denial-of-service attacks within rate limits.

**Privacy Adversary (A_PRIV):** Seeks to correlate photographers to images or track device usage patterns. Manufacturer authorities observe NUC hash fingerprints during validation (e.g., "fingerprint a3f8... validated 50 times in November") but fingerprints are not linked to camera serial numbers or owners in MA databases. Environmental conditions during sensor calibration (temperature, humidity, equipment variance) make regenerating identical fingerprints computationally infeasible—even seizing a camera cannot prove it produced a specific fingerprint. Blockchain observers see image hashes and metadata hashes but cannot determine which cameras created images. Correlation requires compromising manufacturer production records (linking fingerprints to serials during manufacturing) AND MA validation logs AND submission server logs. Assumes A_PRIV cannot break AES-256, SHA-256, or ECDSA.

**Software-Bypass Adversary (A_SW):** This adversary possesses a legitimate, authenticated camera but seeks to hide the use of generative AI tools. They do not attempt to break the hardware encryption; instead, they use a compromised or "patched" version of an editing suite that performs content-aware modifications (Level 2) while falsely reporting them as standard adjustments (Level 1). The goal is to obtain a "Validated" status for a manipulated image.

**Adversary capabilities are bounded:** A_HW cannot extract all devices economically, A_NET cannot exceed Byzantine threshold (33% validators), A_PRIV cannot break standard cryptographic primitives. Adversaries may cooperate but face coordination costs increasing with geographic and organizational distribution.

## 6.3 Security Claims

The system architecture ensures the following security properties:

**Claim 1 (Blockchain Observer Non-Identification):** Given blockchain records containing {image_hash, modification_level, parent_hash, metadata_hashes, server_ids, timestamp}, blockchain observers cannot determine which camera created which image with probability >1/k (where k = anonymity set size, typically thousands of cameras) without compromising both manufacturer authority AND submission servers.

Justification: Blockchain records contain image_hash but no device identifiers. During validation, manufacturers decrypt device identity but never observe corresponding image hashes (not transmitted in validation requests). Submission servers observe image hashes but cannot decrypt device identities.

Correlation requires simultaneously compromising manufacturer key storage AND server logs, exceeding the Byzantine threshold.

**Claim 2 (Manufacturer Non-Correlation):** Manufacturer authorities cannot correlate device validations to blockchain images with probability $>1/k$ (k = anonymity set size, typically thousands) without compromising submission servers.

Justification: Manufacturers decrypt device identities during validation but never observe corresponding image hashes (not transmitted in validation requests). Manufacturers can later observe image_hash values on the public blockchain but cannot link them to prior validation events without submission server logs associating validation timestamps to image submissions. Anonymity sets of thousands of cameras per key bound guessing probability to $1/k$.

**Claim 3 (Manufacturer Device Blindness):** Manufacturer authority cannot identify which specific cameras validated images, even with complete database access. MA database contains NUC hashes without device identifiers, enabling validation without tracking individual equipment.

Justification: MA validation performs only membership queries: "Is this NUC_hash legitimate?" Database structure prevents answering "Which camera validated?" or "How many times was camera X used?" Even MA compromise reveals only population of legitimate devices, not individual usage patterns.

**Claim 4 (Authentication Integrity):** A record R is accepted by blockchain validators if and only if:

- R.image_hash = SHA-256(pixel_data) of image captured by legitimate camera C, AND
- C possesses valid NUC_hash in manufacturer database OR valid PRNU-seeded ECDSA keypair, AND
- Manufacturer Authority validated C via NUC-hash matching and signed approval, AND
- Two independent submission servers $S_i$, $S_j$ ($i \neq j$) verified camera signatures and obtained Manufacturer Authority approvals

See Section 6.4 for detailed attack-defense analysis of each layer.

Justification: Camera signatures prevent packet tampering (submission → server). Dual-approval prevents single-server forgery (server → blockchain). Validator consensus prevents blockchain manipulation (Byzantine tolerance). Three-layer defense requires compromising camera secure element AND two Submission Servers AND validator majority) exceeding Byzantine threshold by significant margin.

**Claim 5 (Single-Server Cannot Forge):** Adversary A_NET controlling single submission server S_i cannot create valid blockchain record without valid manufacturer approval AND second independent submission server S_j approval.

Justification: Blockchain validators enforce dual-approval protocol: require two MA-signed approvals with different server_id fields AND two submission server signatures. Single compromised server S_i receives legitimate submission and obtains approval_1 from MA. However, S_i cannot forge second approval because: (1) approval_2 must have different server_id than approval_1 (validator checks), (2) Manufacturer Authority signatures bind approval to specific server_id, (3) S_i cannot request approval_2 as S_j without S_j credentials, (4) S_i cannot forge S_j's signature on the blockchain record. Server signatures provide additional defense layer beyond MA approval verification. Therefore single-server compromise insufficient for record forgery.

**Claim 6 (Byzantine Fault Tolerance):** System tolerates up to $f = \lfloor(n-1)/3\rfloor$ compromised validators through GRANDPA consensus [12], which requires 67% validator agreement for block finalization. For recommended deployment (n=10 validators), system maintains integrity with up to 3 compromised validators.

Justification: GRANDPA consensus requires 67% validator agreement for block finalization. Compromised validators (≤33%) cannot finalize invalid blocks. Honest validators (≥67%) reject blocks violating protocol rules (missing approvals, invalid signatures, conflicting token_excerpts). Byzantine threshold f < n/3 ensures honest validators always maintain finalization majority.

## 6.4 Attack-Defense Matrix

Systematic analysis of attack vectors and architectural defenses:

**Attack: Identify Device from Blockchain**

Adversary: A_PRIV (blockchain observer)

Defense: Blockchain records contain image hashes and metadata hashes but no device identifiers or encrypted tokens. Device identification requires compromising manufacturer validation logs (to learn device identities from decrypted tokens) AND submission server logs (to correlate validation events to image hashes). Blockchain validators verify MA signatures during submission but discard them after validation; no manufacturer identifiers appear in stored records.

Security Property: Blockchain Observer Non-Identification

Residual Risk: If both manufacturer authority and submission server compromised, correlation possible.

Mitigation: Requires two independent compromises across organizational boundaries.

**Attack: Correlate Device to Image (Manufacturer)**

Adversary: A_PRIV (manufacturer authority)

Defense: AES-256 encrypted tokens with key table anonymity sets (typically thousands of cameras per key). Manufacturer authority decrypts device identity during validation but never observes corresponding image hashes (not transmitted in validation requests). MA can observe image hashes on public blockchain but cannot link them to validation events without submission server logs.

Security Property: Manufacturer Non-Correlation

Residual Risk: If manufacturer authority compromises submission server, gains correlation capability.

Mitigation: Submission servers operated by journalism organizations, not manufacturers.

**Attack: Timing Correlation (Real-Time Blockchain Monitoring)**

Adversary: A_PRIV (blockchain observer attempting device tracking)

Defense: Blockchain timestamps rounded to 10-minute increments (e.g., records from 10:00:00-10:09:59 all show "10:00"). Blockchain observers see only rounded timestamps, not precise submission times. Effective anonymity set equals all records posted globally within the same 10-minute window (typically hundreds for active deployment).

Security Property: Temporal Privacy

Residual Risk: Low-traffic periods (<10 records per window) reduce anonymity sets. High-volume photographers (>100 images per window) may dominate timestamp bins.

Mitigation: Coalition monitors distribution. During low traffic, servers extend windows to 20-30 minutes to maintain minimum anonymity set size.

**Attack: Frequency Analysis (Operational Tempo Tracking)**

Adversary: A_PRIV (manufacturer authority tracking device usage)

Defense: MA logs record fingerprint activity at month-precision (e.g., "fingerprint a3f8... validated 50 times in 2025-11") but fingerprints are not linked to serial numbers or owners. Environmental calibration conditions make fingerprints non-regenerable—cannot determine which physical camera produced which fingerprint. Even with MA database access, adversary learns only that anonymous fingerprints validated images, not which journalists own those cameras.

Security Property: Limited Surveillance Resistance

Residual Risk: Compromising manufacturer production records (linking fingerprints to serials during manufacturing) AND MA validation logs enables device tracking.

Mitigation: Month-precision limits granularity. Production records maintained separately from validation database with different access controls.

**Attack: GPS Location Brute-Force**

Adversary: A_PRIV (privacy adversary with approximate location knowledge)

Defense: HMAC-SHA256(GPS, nonce) truncated to 64 bits (8 bytes). Search space: 2^64 attempts even with known approximate location. Nonce stored in published image EXIF enables verification.

Security Property: Metadata Privacy

Residual Risk: None if nonce remains secret. If nonce leaked, still requires 2^64 brute-force (computationally infeasible).

**Attack: Metadata Cross-Field Brute-Force**

Adversary: A_PRIV (privacy adversary with leaked nonce)

Attack Vector: If same nonce used for all metadata fields and timestamp is brute-forced successfully, adversary could use the discovered nonce to attack other fields (GPS, owner).

Defense: Domain separation via context prefixes: HMAC(nonce, "BM-v1-timestamp:" || timestamp), HMAC(nonce, "BM-v1-geolocation:" || GPS), HMAC(nonce, "BM-v1-owner:" || owner_id). Even if nonce is discovered through timestamp brute-force, it cannot be used to attack other fields due to different context strings.

Security Property: Metadata Privacy

Residual Risk: None - domain separation prevents cross-field attacks cryptographically.

**Attack: Secure Element Extraction and Credential Forgery**

Adversary: A_HW (hardware adversary with device physical access)

Attack Vector: Extract secure element keys via invasive analysis (~$100K per device), then either (1) clone credentials to forged camera hardware, or (2) use extracted keys to sign fraudulent authentications from software.

Defense: Secure element tamper resistance (TPM 2.0, StrongBox, Secure Enclave) makes extraction require electron microscopy and specialized equipment. For manufacturing-calibrated cameras, adversary must also compromise manufacturer database to register forged NUC hash.

Security Property: Authentication Integrity

Residual Risk: Nation-state adversaries may accept $100K cost for high-value target surveillance (single journalist). Cannot scale to spam operations (would require $100K × thousands of devices = tens of millions).

Mitigation: Economic infeasibility at scale ($100K × N devices). Coalition threat intelligence sharing detects coordinated forgery.

**Attack: Manufacturer Authority Private Key Compromise**

Adversary: A_HW (hardware adversary targeting MA infrastructure)

Defense: Multi-layer defense. (1) HSM tamper resistance with physical tamper detection. (2) Dual-approval protocol requires both MA signatures AND submission server signatures; compromising MA alone cannot forge server signatures. (3) Event-driven key rotation enables response to detected compromise. (4) Multi-party key generation via threshold ceremony with coalition oversight.

Security Property: Authentication Integrity (dual-approval defense-in-depth)

Residual Risk: If adversary compromises MA HSM AND two submission servers, can forge records. Requires three independent compromises across organizational boundaries.

Mitigation: Geographic distribution of trust brokers, HSM physical security, event-driven key rotation limits compromise window.

**Attack: Single Server Record Forgery**

Adversary: A_NET (network adversary)

Defense: Dual-approval architecture requires two MA-signed approvals from different submission servers. Each MA signature approves a specific server's validation request and cannot be reused by another server.

Security Property: Byzantine Tolerance (submission infrastructure)

Residual Risk: Two submission servers compromised enables forgery. Mitigation: Servers operated by different trust brokers in different jurisdictions.

**Attack: Content Injection via Malicious ISP Module**

Adversary: A_HW (compromised camera firmware)

Attack Vector: Adversary injects content modifications (object removal, generative fill) into camera's ISP pipeline, attempting to pass content-altered images as standard ISP processing.

Defense: Software Submission Authority validates by replaying reported operations on 100 random 64×64 patches from parent image. Rejects if cumulative deviation exceeds manufacturer threshold (15-18%) or mismatches reported score by >5%.

Security Property: ISP Processing Integrity

Residual Risk: Sampling-based validation could miss targeted edits in non-sampled regions. Random patch selection makes evasion difficult (requires modifying >85% of image to maintain low sampled deviation).

Mitigation: Random patch selection, manufacturer-specific thresholds, manual review.

**Attack: Replay Valid Authentication**

Adversary: A_NET (network adversary with intercepted packets)

Defense: record_hash cryptographically binds certificate to specific image_hash. Replaying certificate with different image fails validation (server recomputes hash of new record, doesn't match hash in certificate). Blockchain prevents duplicate image_hash records (each image authenticated once).

Security Property: Authentication Integrity (prevents replay)

Residual Risk: None. Cryptographic binding and blockchain uniqueness constraint prevent replay attacks.

**Attack: Validator Collusion (>33%)**

Adversary: A_NET (network adversary)

Defense: Byzantine consensus (GRANDPA) requires 67% validator agreement. Geographic distribution across trust brokers in multiple jurisdictions increases coordination cost. Mission-aligned governance reduces incentives for collusion.

Security Property: Byzantine Fault Tolerance

Residual Risk: If >33% validators compromised, adversary can finalize invalid blocks. This exceeds the Byzantine fault tolerance threshold; security properties fail by design (assumption violation, not mitigatable).

## 6.5 Proof Sketches

We provide intuition for key privacy and integrity claims showing novel architectural properties. Full formal verification remains future work (see Section 8).

**Proof Sketch for Claim 1 (Blockchain Observer Non-Identification):**

Given: Blockchain observer O sees record R = {image_hash, modification_level, parent_hash, metadata_hashes, server_ids, timestamp} on public blockchain.

Goal: Show O cannot identify which camera created image_hash with probability >1/k where k = global camera population (millions).

Step 1: Observer O extracts image_hash H from blockchain record R.

Step 2: To identify the camera, O needs to learn which physical camera created H.

Step 3: Camera identification requires three pieces of information:

(a) NUC_hash ↔ encrypted_token mapping (MA validation logs)
(b) Encrypted_token ↔ image_hash association (submission server logs)
(c) NUC_hash ↔ camera_serial mapping (manufacturer production records, separate from MA database)

Analysis:

MA sees: NUC_hash (from decrypting tokens) but NOT image_hash (not transmitted) or camera serials (not in MA database). MA logs: {timestamp, validation_result} only.

Submission server sees: image_hash but NOT NUC_hash (cannot decrypt tokens without MA keys).

Blockchain records: image_hash only—no NUC_hash, encrypted_token, or device identifiers.

Attack requires simultaneous compromise of:

1. MA validation logs (obtain NUC_hash)
2. Submission server logs (link encrypted_token ↔ image_hash)
3. Manufacturer production records (link NUC_hash ↔ camera_serial)

Without all three:

- Only (a): Knows which NUC_hashes validated, but not which images or physical cameras
- Only (b): Knows images authenticated, but not by which NUC_hashes or cameras
- (a) + (b): Can link NUC_hashes to images, but NUC_hashes remain anonymous—cannot identify physical cameras without production records

Conclusion: Requires three independent compromises across organizational boundaries (MA validator, submission server, manufacturer production). Even with key table knowledge (k ≈ thousands), guessing probability without production records: $1/k$.

Probability of identification without triple compromise: $\leq 1/N$ (global population) or $1/k$ (with key table knowledge).

**Proof Sketch for Claim 2 (Manufacturer Non-Correlation):**

Given: MA observes device validation events and can observe image hashes on public blockchain.

Goal: Show MA cannot correlate device to image with probability $>1/k$ where $k$ = anonymity set size (typically thousands).

Step 1: Camera sends {image, manufacturer_cert, camera_signature} to submission server S.

Step 2: S forwards manufacturer_cert (containing encrypted_token E, key_reference, deviation_data) to MA for validation. Image_hash is NOT included in validation request.

Step 3: MA decrypts E using key from table → obtains NUC_hash.

Step 4: MA performs membership check: Is NUC_hash in legitimate device set?

Step 5: MA returns signed approval (if hash found) or rejection (if not found). MA does not log device identifiers—only validation timestamp and result.

Analysis: MA confirms "this is a legitimate camera" without learning "this is camera serial #ABC123." Even with full MA database access, adversary cannot identify which cameras validated images or track individual device usage patterns.

Submission server S observed {image_hash, encrypted_token} association during step 1 but cannot decrypt E (lacks key table). Therefore: - MA: can decrypt tokens → learns device_ID, cannot associate to image_hash - S: can associate image to token, cannot decrypt → doesn't learn device_ID

Correlation requires both capabilities. MA must compromise S to obtain association logs, or S must compromise MA to obtain decryption keys. Both exceed single-party adversary capability.

Probability of MA correctly guessing device_ID ↔ H correlation without compromising S: $1/k$ where $k$ = anonymity set size (typically thousands).

**Proof Sketch for Claim 5 (Single-Server Cannot Forge):**

Given: Adversary controls submission server S_1.

Goal: Show S_1 cannot create valid blockchain record without second server S_2.

Step 1: Camera submits legitimate authentication {image_hash, manufacturer_cert, camera_signature} to S_1.

Step 2: S_1 forwards manufacturer_cert to MA and receives signed approval_1 (MA signature over record_hash with server_id = S_1).

Step 3: Adversary attempts to create fraudulent blockchain record using only approval_1.

- Blockchain validation requirements:
- Two MA-signed approvals: approval_1, approval_2
- Both approvals must sign same record_hash
- approval_1.server_id ≠ approval_2.server_id (different servers)
- Two server signatures from those servers

Adversary options:

Option A: Reuse approval_1 as approval_2 Result: REJECTED (blockchain validators check approval_1.server_id ≠ approval_2.server_id, fails same-server test)

Option B: Request approval_2 from MA claiming to be S_2 Result: REJECTED (MA verifies server credentials via TLS certificate, S_1 lacks S_2's private key)

Option C: Forge MA signature on approval_2 Result: REJECTED (blockchain validators verify MA signature using MA_public_key, forgery detected)

Option D: Forge server_2 signature on blockchain submission Result: REJECTED (blockchain validators verify both server signatures, S_1 lacks S_2's private key)

Option E: Compromise second server S_2 and obtain legitimate approval_2

Result: SUCCESS but requires TWO server compromises (exceeds single-server adversary assumption)

Conclusion: Single compromised server cannot create valid blockchain record. Two independent server compromises required. With servers operated by different journalism organizations in different jurisdictions, this requires coordinating attacks across organizational boundaries.

## 6.6 Computational Security Assumptions

Security claims hold under standard cryptographic assumptions:

**AES-256 Security:** No adversary can distinguish AES-256(device_ID, MA_key) from random with non-negligible probability. Specifically, advantage $AdvAES(A) < 2^{-128}$ for all probabilistic polynomial-time adversaries A.

**SHA-256 Collision Resistance:** Finding distinct images I, I' where SHA-256(I) = SHA-256(I') is computationally infeasible. Specifically, finding collisions requires $O(2^{128})$ hash evaluations (birthday bound).

**ECDSA Signature Security:** Forging ECDSA signatures without private key is computationally infeasible under discrete logarithm hardness assumption. Specifically, forging signatures requires solving discrete logarithm problem over elliptic curve group (complexity: $O(2^{128})$ for secp256k1).

**HMAC-SHA256 Security:** HMAC-SHA256(GPS, nonce) is indistinguishable from random for adversaries without knowledge of nonce. Even with approximate location knowledge, 64-bit truncation provides $2^{64}$ search space (computationally infeasible for real-time attacks).

**Byzantine Threshold Assumption:** At most $f < n/3$ validators are compromised where n = total validator count. System designed for $n \geq 4$ (minimum) to $n = 10$ (recommended deployment). With n=10, tolerates f=3 compromised validators.

**Anonymity Set Size Assumption:** Key tables contain $k \geq 1{,}000$ devices per authentication path (typical deployment: $k \geq 1{,}000$). Privacy properties degrade if $k < 100$ (correlation probability becomes non-negligible).

These assumptions are standard in applied cryptography and represent conservative security margins. AES-256, SHA-256, and ECDSA have withstood decades of cryptanalysis without practical breaks.

## 6.7 Known Limitations

We acknowledge security limitations and residual risks:

**Limitation 1 - Manufacturer Database Compromise (Limited Exposure):**

Manufacturer authority maintains database of NUC hashes from legitimate cameras, enabling validation without storing device identifiers. MA observes aggregate validation volume (e.g., "500 validations in November 2025") but cannot track individual cameras.

Even under complete MA compromise, adversary learns only:

- Population of legitimate cameras (NUC hash set)
- Aggregate validation frequency (monthly totals)

Adversary does NOT learn:

- Which specific cameras validated images
- Individual camera usage patterns
- Camera serial numbers or owner identities

Mitigation: MA logs retained for maximum 3 years. Validation queries are stateless (membership check only). Coalition audits verify MA does not log device identifiers.

Residual Risk: If adversary compromises BOTH manufacturer production records (linking NUC hashes to serial numbers during manufacturing) AND MA validation logs, correlation becomes possible. This requires compromise of separate systems (factory vs. validator) with different access controls.

Assessment: Removing device identifiers from MA database significantly reduces surveillance risk. Even insider access to MA validation infrastructure reveals only that authentication system is being used, not which journalists or cameras are active.

**Limitation 2 - Validator Collusion (Byzantine Assumption Violation):**

If adversary compromises >33% of validators, system security fails: malicious validators can finalize invalid blocks, censor legitimate submissions, or selectively block photographers. Byzantine consensus (GRANDPA) provides security only under $f < n/3$ malicious validator assumption.

Mitigation: Geographic distribution across trust brokers in multiple jurisdictions increases coordination cost for collusion. Mission-aligned governance (journalism organizations, fact-checking networks, press freedom advocates) reduces incentives versus commercial motivations. Validator admission requires 67% approval, enabling coalition to reject suspect organizations.

Residual Risk: Human trustworthiness and governance depend on institutional integrity of trust brokers. No technical mechanism prevents coordinated human collusion across organizational boundaries.

**Limitation 3 - Secure Element Extraction (High-Value Target Surveillance):**

Adversaries with specialized laboratory equipment can extract secrets from camera secure elements through invasive techniques [19,34,35]. While Common Criteria EAL5+ certification [18] provides strong protection, successful attacks require clean room facilities, electron microscopy, and expert personnel. Extraction affects only the single compromised device.

Mitigation: Economic and technical barriers prevent mass exploitation. Infrastructure requirements are inconsistent with spam-scale operations requiring thousands of devices. Coalition threat intelligence sharing enables detection of coordinated forgery. No technical mitigation exists for determined adversaries targeting individual high-value devices; detection relies on validation pattern monitoring and device blacklisting.

Residual Risk: No technical mitigation exists for determined adversaries willing to invest resources in targeting individual high-value devices. Detection relies on validation pattern monitoring and device blacklisting upon detection.

**Limitation 4 - Formal Verification Gap:**

We provide security analysis using applied π-calculus and ProVerif verification for core privacy properties (Appendix A), demonstrating that architectural separation prevents manufacturer and blockchain observer correlation. Additional properties are analyzed via proof sketches showing how cryptographic primitives and protocol structure enforce security goals. Complete mechanized formal verification across all security properties using proof assistants (Coq, Isabelle) remains future work.

Scope of Current Verification: ProVerif establishes observational equivalence for privacy properties under Dolev-Yao adversary model. Proof sketches demonstrate authentication integrity, Byzantine fault tolerance, and anti-censorship properties through architectural analysis. Computational security assumptions (AES-256, SHA-256, ECDSA) rely on decades of cryptanalytic research rather than mechanized proof.

Industry Context: Our verification approach follows standard practice for deployed cryptographic systems. TLS, Signal Protocol, and WireGuard similarly combine mechanized proof for critical protocol properties with rigorous argument for secondary properties, rather than complete formal verification. Full mechanized verification of all properties would require significant research investment beyond individual project scope.

Assessment: The combination of ProVerif verification for privacy properties and detailed proof sketches for other security goals provides appropriate confidence for a novel authentication architecture. Gaps in mechanized verification represent acknowledged research directions rather than deployment blockers, consistent with how production security systems are validated before widespread deployment.

**Limitation 5 - Side-Channel Attacks (Out of Scope):**

This analysis focuses on cryptographic protocol security and does not address side-channel attacks (timing analysis, power consumption, electromagnetic emanation). Secure element implementations may be vulnerable to side-channel extraction techniques beyond cryptographic security model.

Mitigation: Rely on secure element manufacturers (STMicroelectronics, NXP, Infineon) to implement side-channel countermeasures (constant-time operations, power analysis resistance, shielding). Trust broker coalition can specify secure element requirements (Common Criteria EAL5+, FIPS 140-3 Level 3) for manufacturer certification.

Assessment: Side-channel attacks require physical access and sophisticated equipment ($10K-100K). Relevant for high-value target surveillance but not mass spam. System design assumes secure element physical security properties hold per manufacturer specifications.

**Limitation 6 - ISP Processing Integrity (Heuristic Defense):**

ISP deviation detection monitors for firmware-level manipulation but provides heuristic monitoring rather than cryptographic guarantees. Rigorous evaluation requires manufacturer ISP implementations, editing software internals, and large labeled datasets—resources unavailable to individual researchers. Proper validation is deferred to funded phases with industry partnerships. We treat this as an evolving adversarial detection layer, not a foundational trust mechanism.

Trust Assumption: We assume camera manufacturers implement honest ISP pipelines and protect firmware integrity through signing and secure boot. Adversaries with supply chain access or firmware signing key compromise could inject content modifications while reporting false deviation scores.

Mitigation: Manufacturer firmware signing prevents unauthorized firmware. Cameras with known vulnerabilities can be revoked. ISP processing thresholds (deviation <15-18%) bound possible manipulation. Coalition threat intelligence enables detection of coordinated attacks. However, no technical mitigation exists for determined adversaries compromising manufacturer firmware infrastructure—detection relies on pattern monitoring and device blacklisting.

Assessment: Removing ISP deviation entirely would not compromise core authentication (NUC fingerprints, manufacturer validation, blockchain integrity). This layer provides defense-in-depth against a specific attack vector but is explicitly scoped as advisory monitoring rather than cryptographic security.

# 7 IMPLEMENTATION AND PERFORMANCE

Prototype implementation validates feasibility using Raspberry Pi 4 with HQ Camera (Sony IMX477 sensor). Complete authentication pipeline demonstrated from capture through blockchain verification.

**Camera Overhead:** Parallel hashing during capture adds <100ms latency (SHA-256 computation, signature generation). Negligible impact on user experience. Memory: 8KB secure element storage for keys and NUC hashes.

**Network Requirements:** 2KB submission packet (image hash, encrypted token, certificate, signature). Standard HTTPS transport. Graceful degradation: queues submissions if offline, retries with exponential backoff.

**Blockchain Performance:** 150 bytes per record. At 1 million images/day: 55GB/year growth. Query latency: <50ms for hash lookup via indexed database. Validator requirements: commodity hardware (4-core CPU, 16GB RAM, 1TB SSD) handles projected load.

**Economic Model:** Validator node operating costs ~$100-150/month (storage, bandwidth). Camera hardware integration adds ~$2 per device (secure element chip at manufacturing volume). Economic spam barrier: forgery requires legitimate manufacturer-validated cameras; mass acquisition detectable.

Performance targets are based on cryptographic operation benchmarks for ARM Cortex-A processors [37], typical database indexed query performance [38], and network round-trip latency statistics [39].

# 8 DISCUSSION

This section examines the relationship between the Birthmark Standard and existing authentication approaches, deployment strategies for practical adoption, and future extensions of the system.

## 8.1 Relationship to C2PA

The Birthmark Standard complements C2PA rather than replacing it. Both address photographic authentication but differ in technical approach and resulting properties.

C2PA embeds cryptographically signed metadata directly into image files, providing rich provenance (full editing history, multiple signers, equipment details) when metadata preserved. However, social media platforms strip embedded credentials through compression and lossless transformations (affecting the vast majority of distributed images. C2PA verification infrastructure is operated by commercial platforms, creating dependency on continued corporate support for validation services.

Birthmark stores authentication records on consortium blockchain independent of image files. Verification depends only on pixel hash, surviving all metadata loss and format conversion. Governance by mission-aligned trust brokers rather than commercial platforms.

**Complementary deployment:** C2PA optimal for professional workflows where metadata preserved (newsroom editing, archival systems, legal evidence). Birthmark optimal for public distribution where metadata stripped (social media, viral content, adversarial contexts). Organizations can deploy both: C2PA for complete provenance in controlled environments, Birthmark for last-resort verification when metadata lost.

## 8.2 Deployment Strategy

PRNU-seeded keypair architecture (Section 4.2) enables deployment without camera manufacturer cooperation. Smartphone apps use existing secure hardware (Android StrongBox, iOS Secure Enclave), eliminating firmware partnership dependency. This inverts traditional standardization requiring manufacturer buy-in before deployment.

**Early Adoption Pathways**

Deployment prioritizes contexts where (1) authentication fraud is visible and immediate, (2) submission/validation processes are controlled, and (3) early adopters have clear incentives. We identify two promising markets based on documented fraud concerns and controlled validation environments:

**E-commerce copyright enforcement** provides immediate deployment traction. Independent sellers face persistent image theft—product photos stolen for counterfeit listings, artwork reproduced without authorization, design work used by unauthorized retailers. Current DMCA processes require proving copyright ownership; authenticated images provide cryptographic proof, accelerating platform takedowns. Business owners experiencing revenue loss from theft represent motivated early adopters with clear ROI, unlike speculative adoption for theoretical future threats. AI-generated image submissions to major photography competitions have been documented since 2023 [25,26,27], creating reputational risk for competition organizers.

**Phase 2 market validation:** Direct outreach to 50+ independent merchants and photographers experiencing image theft, surveying willingness to adopt authentication tools. Beta testing with 10-20 sellers and 3-5 photography competitions. Success metrics: 50%+ of surveyed participants indicate willingness to adopt within 12 months; beta testers report successful fraud reduction.

**Photography competitions** provide parallel pathway with different incentives: reputation protection against AI submission fraud rather than revenue protection against theft. Competitions already control both submission requirements and judging, enabling straightforward integration of authentication requirements. Competition winners are often working photojournalists, creating professional adoption pipeline. Success across both commercial (e-commerce) and reputational (competitions) contexts demonstrates broad applicability, attracting manufacturer partnerships and journalism organization adoption.

**Platform Verification Integration**

Platform verification requires minimal integration: browser extensions query blockchain directly via JSON-RPC without platform cooperation. Social media platforms may integrate verification badges optionally, but system succeeds through browser-based verification. Users install extension once; verification happens automatically while browsing.

**Coalition Governance Evolution**

Governance evolves progressively: Foundation operates initial infrastructure, expands to working group with early partners, transitions to formal consortium with trust broker voting rights. Blockchain supports runtime upgrades via on-chain governance, enabling protocol evolution without hard forks or coordinated validator downtime. This eliminates the operational burden where traditional blockchains require significant coordination overhead per upgrade.

**Operational Monitoring**

Coalition deploys automated monitoring for validator health (uptime, validation patterns, software versions). Anomaly detection triggers manual review and potential node removal via governance vote. Continuous monitoring ensures Network reliability and detects potential compromise attempts early.

# 9 CONCLUSION

We present the Birthmark Standard, an open authentication architecture establishing verifiable photographic provenance while maintaining strong privacy guarantees and decentralized governance.

The system addresses technical limitations and structural risks in existing metadata-based approaches through hardware roots of trust, consortium blockchain storage, and privacy-preserving validation protocols that defend against surveillance while enabling verification.

**The architecture makes infrastructure control explicit:** who operates the systems determining photographic authenticity? Existing solutions depend on corporate platforms with commercial incentives potentially diverging from press freedom. When Reuters pulled AI-generated images from their wire service, they relied on Adobe's infrastructure for detection. This dependency creates structural fragility. We cannot predict Microsoft's or Adobe's policies in 5-10 years, but we can predict that CPJ, RSF, and IFCN will prioritize press freedom and credible information - their founding missions.

**Technical validation demonstrates production readiness:** <100ms camera overhead, <500ms verification latency, architecture scaling beyond millions of daily authentications with $1.2K-1.8K annual node costs accessible to trust brokers. The complete implementation from sensor capture through public verification has been validated on commodity hardware.

This work enables credible photojournalism when visual evidence matters most: protests, conflicts, disasters, humanitarian crises. News organizations can differentiate through consistent authentication. Viewers can distinguish genuine photographic evidence from AI-generated content.

The Birthmark Standard offers public infrastructure controlled by trust brokers, hardware roots of trust surviving metadata loss, privacy preservation preventing surveillance, and open governance resisting monopolization. We build this because press freedom requires independence, provenance must survive distribution, authentication cannot enable surveillance, and verification infrastructure should serve public interest rather than commercial incentives.

If deployment faces obstacles, our research remains public as defensive prior art preventing patent enclosure. However, our prototype validation demonstrates technical feasibility; the remaining challenges are organizational (coalition formation) rather than technical.

Complete technical specifications, reference implementations, and governance frameworks are available under Apache 2.0 license at github.com/Birthmark-Standard/Birthmark.

## 9.1 Future Work

The authentication architecture extends beyond photography to broader media verification challenges:

- **Video authentication:** Frame-by-frame hashing with temporal linking creates verifiable video chains. Each frame receives blockchain record linked to previous frame via parent_hash, establishing continuous custody from capture through editing. Detecting frame insertion, deletion, or AI-generated segments becomes blockchain query operation.

- **Document authentication:** Content hashing establishes immutable provenance for legal documents (wills, deeds, contracts), court records (opinions, statutes), and notarized materials. Blockchain timestamping proves document existence at specific dates; verification prevents AI citation hallucination (documented failures where AI legal assistants cite non-existent case law) and establishes tamper-evident records for legal proceedings.

- **Screenshot and derivative authentication:** Platform partnerships enable automatic verification of screenshots containing authenticated images. Image identification extracts embedded content, queries blockchain for original records, creates linked records (modification_level=2) with parent_hash preserving provenance while accounting for unknown surrounding context.

- **Platform integration:** Social platforms (Pinterest, Bluesky) can automatically submit derivative records when authenticated images are uploaded—crops, filters, and edits link to originals via parent_hash, maintaining verifiable custody chains through viral distribution. Users verify via browser extensions; platforms optionally display verification badges.

[38] PostgreSQL Global Development Group, "PostgreSQL 14 Documentation: Indexes and Performance," https://www.postgresql.org/docs/14/indexes.html, 2021.

[39] Cloudflare, "Network Performance Update: Q4 2024," Cloudflare Radar Insights, https://radar.cloudflare.com/reports/speed-and-reliability, 2024.27

# Appendix A: Formal Privacy Analysis via Applied π-Calculus

This appendix provides a formal verification of the Birthmark Standard's privacy-preserving properties using the applied π-calculus and the ProVerif verification tool. We establish that the architectural separation between image content validation and device identity verification prevents correlation attacks by both the Manufacturer Authority and external blockchain observers.

## A.1 Threat Model and Trust Assumptions

**Adversary Model.** We assume a Dolev-Yao network attacker with full control over all public communication channels. The adversary can intercept, replay, and compose messages but cannot break symbolic cryptographic primitives (decrypt without keys, forge signatures without signing keys, or find hash collisions).

**Trust Assumptions.** We assume non-collusion between the Submission Server $S$ and Manufacturer Authority $M$. Each entity operates independently and does not share internal state. The blockchain $B$ is modeled as a public append-only channel accessible to all observers, including the attacker.

**Cryptographic Idealization.** Under the symbolic model, we treat all cryptographic operations as perfect black-box constructors:

- $\text{aenc}(m, pk)$ / $\text{adec}(c, sk)$: Asymmetric encryption/decryption where $\text{adec}(\text{aenc}(m, pk), sk) = m$ iff $(pk, sk)$ is a valid keypair.
- $\text{hash}(m)$: One-way collision-resistant hash with no inverse.
- $\text{sign}(m, sk)$ / $\text{verify}(\sigma, m, pk)$: Digital signatures where $\text{verify}(\text{sign}(m, sk), m, pk) = \text{true}$ iff $(pk, sk)$ is a valid keypair.

**Manufacturer Registry.** The Manufacturer Authority holds a table $R = \{T_1, \ldots, T_N\}$ of $N$ valid device tokens, modeling the anonymity set. Each camera $C_i$ possesses a unique token $T_i \in R$ and the Manufacturer's public key $PK_m$.

## A.2 Protocol Specification

We model the protocol as four concurrent processes: Camera, Submission Server, Manufacturer Authority, and Blockchain.

**Camera Process** $C(T, PK_m, img)$:

```
C(T, PK_m, img) :=
  let H = hash(img) in
  let cipher = aenc(T, PK_m) in
  out(pub, (H, cipher))
```

The camera computes the image hash $H$ and encrypts its device token $T$ under the Manufacturer's public key, then publishes $(H, \text{aenc}(T, PK_m))$ on the public channel.

**Submission Server Process** $S$:

```
S :=
  in(pub, (h:bitstring, enc_token:bitstring));
  out(mfr_channel, enc_token);
  in(mfr_channel, decision:bool);
  if decision = true then out(blockchain, h)
```

The server receives the tuple $(H, \text{aenc}(T, PK_m))$, forwards only the encrypted token to the Manufacturer Authority via a private channel mfr_channel, receives a validation decision, and conditionally publishes $H$ to the blockchain. Critically, $S$ never decrypts the token and $M$ never receives $H$.

**Manufacturer Authority Process** $M(SK_m, R)$:

```
M(SK_m, R) :=
  in(mfr_channel, enc_token:bitstring);
  let T = adec(enc_token, SK_m) in
  let decision = (T ∈ R) in
  out(mfr_channel, decision)
```

The Manufacturer decrypts the token using its private key $SK_m$, checks membership in the registry $R$, and returns a boolean validation result. The Manufacturer's internal state never contains any image hash.

**Blockchain Process** $B$:

```
B := in(blockchain, h:bitstring); out(public_obs, h); B
```

The blockchain is modeled as a public output channel where any observer (including the attacker) can read published hashes.

### A.3 Formal Privacy Properties

We establish two privacy properties via observational equivalence in the applied π-calculus.

**Property A — Manufacturer Non-Correlation.**
*The Manufacturer Authority cannot distinguish which image hash corresponds to a validated device token, even when performing successful validation.*

**Formal Statement:** For any two distinct images $\text{img}_1, \text{img}_2$ and valid token $T \in R$, the following processes are observationally equivalent from $M$'s perspective:

$$C(T, PK_m, \text{img}_1) \mathbin{/\!/} S \mathbin{/\!/} M(SK_m, R) \mathbin{/\!/} B \approx C(T, PK_m, \text{img}_2) \mathbin{/\!/} S \mathbin{/\!/} M(SK_m, R) \mathbin{/\!/} B$$

**Proof Sketch:**
The Manufacturer's view consists only of the sequence of encrypted tokens received on mfr_channel and its own outputs. Since $M$ never receives $H = \text{hash}(\text{img})$, it cannot construct any test that distinguishes $\text{img}_1$ from $\text{img}_2$. The only information flow to $M$ is $\text{aenc}(T, PK_m)$, which is identical in both scenarios.

**Property B — Blockchain Observer Non-Identification.**
*A public observer monitoring the blockchain cannot link a published hash $H$ to a specific camera identity $T$.*

**Formal Statement:** For any observer $O$ with access to the public blockchain channel and the Dolev-Yao network, and for any two cameras $C_1(T_1, PK_m, \text{img})$ and $C_2(T_2, PK_m, \text{img}')$ where $\text{hash}(\text{img}) = \text{hash}(\text{img}') = H$:

$$C_1 \mathbin{/\!/} S \mathbin{/\!/} M \mathbin{/\!/} B \approx_O C_2 \mathbin{/\!/} S \mathbin{/\!/} M \mathbin{/\!/} B$$

**Proof Sketch:** The blockchain contains only $H$, which is a one-way function of image content. Device tokens $T_i$ appear only as $\text{aenc}(T_i, PK_m)$ on the public channel. Since $O$ does not possess $SK_m$, it cannot decrypt these tokens. Furthermore, the submission server's architectural separation ensures $H$ and $T$ are never transmitted together in a form accessible to $O$. Therefore, $O$ cannot construct a test distinguishing which camera produced a given hash beyond the anonymity set size $N = |R|$.

### A.4 ProVerif Model

The following ProVerif code verifies Property A (Manufacturer non-correlation):

```
(* Executable ProVerif model verifying Property A *)

(* --- Types and Functions --- *)
type key.
type skey.
type pkey.

fun pk(skey): pkey.
fun aenc(bitstring, pkey): bitstring.
reduc forall m:bitstring, sk:skey; adec(aenc(m, pk(sk)), sk) = m.
fun hash(bitstring): bitstring.

(* --- Channels --- *)
free pub: channel.
free mfr_private: channel [private].
free blockchain: channel.

(* --- Constants for Logic --- *)
free v_ok: bitstring.
free v_no: bitstring.

(* --- Identities & Keys --- *)
free sk_m: skey [private].
(* FIXED LINE 22: Added letfun to clarify this is a function return, not a process *)
letfun pk_m = pk(sk_m).

free T1: bitstring [private].
free T2: bitstring [private].
free imgA: bitstring [private].
free imgB: bitstring [private].

(* --- Roles --- *)

let camera(token: bitstring, image: bitstring) =
    let h = hash(image) in
    let enc_token = aenc(token, pk_m) in
    out(pub, (h, enc_token)).

let server =
    in(pub, (h: bitstring, enc_token: bitstring));
    out(mfr_private, enc_token);
    in(mfr_private, decision: bitstring);
    if decision = v_ok then out(blockchain, h).

let manufacturer =
    in(mfr_private, enc_token: bitstring);
```

```
        let token = adec(enc_token, sk_m) in
        if (token = T1 || token = T2) then
            out(mfr_private, v_ok)
        else
            out(mfr_private, v_no).

(* --- Verification Execution --- *)

process
    (* We must call the function pk_m inside an 'out' or 'let' here *)
    out(pub, pk_m);
    (
        (!camera(T1, choice[imgA, imgB])) |
        (!server) |
        (!manufacturer)
    )
```

**Verification** **Result.**

ProVerif successfully establishes observational equivalence with the following output:

```
RESULT Observational equivalence is true.
--------------------------------------------------------------
Verification summary:
Observational equivalence is true.
--------------------------------------------------------------
```

This confirms that an attacker with full network observation capabilities (Dolev-Yao model) cannot distinguish between scenarios where the same camera token is used with different images, formally validating Property A. The termination warnings in the full output indicate selection heuristics used by ProVerif's resolution engine and do not affect the validity of the result.

### A.5 Scope and Limitations

This formal analysis establishes privacy properties under symbolic cryptographic assumptions. The following are explicitly out of scope:

1. **Cryptanalytic Hardness.** We do not analyze bit-security of encryption schemes, discrete logarithm hardness, or computational assumptions. The symbolic model treats primitives as idealized black boxes.

2. **Byzantine Consensus.** Blockchain consistency and liveness properties are not verified. We assume $B$ correctly appends validated hashes but do not model validator Byzantine fault tolerance.

3. **Side-Channel Attacks.** Hardware-level information leakage (timing, power analysis, secure element extraction) is not modeled. We assume trusted execution environments operate as specified.

4. **Staged Photo Attacks.** Physical attacks where adversaries capture legitimate images with compromised cameras, then present them as evidence of different events, are sociological rather than cryptographic concerns.

5. **Key Management.** Distribution of $PK_m$ to cameras and secure storage of $SK_m$ at the Manufacturer are assumed correct. Key compromise is not modeled.

6. **Heuristic Accuracy.** The integrity audit described in Section 5.4 is an advisory heuristic, not a cryptographic proof. The reported deviation thresholds (e.g., 15-18%) are empirical benchmarks intended for initial system tuning. While an A_SW might theoretically develop high-precision manipulation tools to mimic standard ISP noise, the non-deterministic nature of the audit sampling is designed to raise the economic and technical cost of such attacks while significantly reducing viable outputs. The failure of a heuristic audit does not invalidate the underlying hardware-rooted proof of sensor authenticity.

This analysis establishes that, given the architectural separation of validation flows and standard symbolic cryptographic assumptions, the protocol achieves its privacy goals: the Manufacturer cannot correlate device identities to image content, and blockchain observers cannot de-anonymize cameras from published hashes beyond the registry size $N$.